\documentclass[12pt]{article}
\begin{document}
\title{The Ultra Relativistic Maxwell and Proca Equations}
\author{C.V. Aditya \& B.G. Sidharth \\
International Institute for Applicable Mathematics \& Information Sciences\\
Hyderabad (India) \& Udine (Italy)\\
B.M. Birla Science Centre, Adarsh Nagar, Hyderabad - 500 063
(India)}
\date{}
\maketitle
\begin{abstract}
We consider the Proca equation which is the Maxwell equation of
electromagnetism for a massive particle, in the ultra relativistic
limit using Snyder-Sidharth Hamiltonian. There is now an extra
parity non-conserving term and we investigate the consequence both
for the Proca equation and the Maxwell equation.
\end{abstract}
\section{Introduction}
We know that Einstein's relativistic energy momentum relation [13]
is given by
\begin{equation}
p_\mu p^\mu = \frac{E^2}{c^2} - p^2 = m^2c^2\label{1}
\end{equation}
or, equivalently
\begin{equation}
E^2 = p^2 c^2 + m^2c^4\label{2}
\end{equation}
Now if we take natural units i.e. $\hbar = c = 1$ we can rewrite
this equation as
\begin{equation}
E^2 = p^2 + m^2\label{3}
\end{equation}
This equation (\ref{3}) can be considered only in the case where
space time is continuous. But if we consider non-commutative
geometry and the Snyder relation of position and momentum [2, 3]
\begin{equation}
[x,p] = \hbar = \hbar [1 + (l/\hbar )^2 p^2]\label{4}
\end{equation}
where $l$ is the minimum length, we can see that if $l \to 0$ we get
back the usual Heisenberg relation of position and momentum.
Substitute equation (\ref{4}) in (\ref{3}) we get
\begin{equation}
E = (m^2 + p^2[1 + l^2 p^2]^{-2})^{1/2}\label{5}
\end{equation}
or,
\begin{equation}
E^2 = p^2 + m^2 - \alpha l^2 p^4\label{6}
\end{equation}
This equation is called Snyder-Sidharth Hamiltonian for bosons and
for fermions it will be more generally [2]
\begin{equation}
E^2 = p^2 + m^2 + \alpha l^2 p^4\label{7}
\end{equation}
This relation becomes important at high energies, for example at
energies which are expected in the LHC (Geneva) [5].
\section{Maxwell Equations}
Let us consider covariant Maxwell's equations [13] given below
\begin{equation}
\partial_\mu F^{\mu \nu} = \frac{4 \pi}{c} j^\nu\label{8}
\end{equation}
This particular equation is in an abridged form with $\mu = 0,1,2,3$
and $\nu = 0,1,2,3$.\\
Adding $\bar{m}^2 A^\nu$ on both side of equation (\ref{8}) we get
\begin{equation}
\partial_\mu F^{\mu \nu} + \bar{m}^2 A^\nu = \frac{4\pi}{c} j^\nu +
\bar{m}^2 A^\nu\label{9}
\end{equation}
\begin{equation}
\partial_\mu (\partial^\mu A^\nu - \partial^\nu A^\mu) + \bar{m}^2
A^\nu = \frac{4\pi}{c} j^\nu + \bar{m}^2 A^\nu\label{10}
\end{equation}
\begin{equation}
\partial_\mu \partial^\mu A^\nu - \partial_\mu \partial^\nu A^\mu +
\bar{m}^2 A^\nu = \frac{4\pi}{c} j^\nu + \bar{m}^2 A^\nu\label{11}
\end{equation}
\begin{equation}
(\partial_\mu \partial^\mu + \bar{m}^2) A^\nu - \partial_\mu
\partial^\nu A^\mu = \frac{4\pi}{c} j^\nu + \bar{m}^2
A^\nu\label{12}
\end{equation}
\begin{equation}
(\partial_\mu \partial^\mu + \bar{m}^2) A^\nu = \frac{4\pi}{c} j^\nu
+ \bar{m}^2 A^\nu - \partial_\mu \partial^\nu A^\mu\label{13}
\end{equation}
We will now introduce the term $\bar{m}^2 = m^2 + \alpha l^2 p^4$ in
equation (\ref{13}) which means that in the environs  of Compton
wavelength the term $\alpha l^2 p^4$ which is going to dominate, is
included and we will see how Maxwell's equations gets modified.\\
From (\ref{13}) we get,
\begin{equation}
(\partial_\mu \partial^\mu + m^2 + \alpha l^2 p^4) A^\nu =
\frac{4\pi}{c} j^\nu + \beta A^\nu - \partial_\mu \partial^\nu
A^\mu\label{14}
\end{equation}
where $\beta = m^2 + \alpha l^2 p^4$.\\
Let us for convenience take momentum space by introducing $p_\mu \to
\imath \hbar \partial_\mu$. This will help us to derive the modified
Einstein's energy momentum relationship.\\
So from equation (\ref{14}) we, get
\begin{equation}
(p_\mu p^\mu + m^2 + \alpha l^2 p^4) A^\nu = \frac{4\pi}{c} j^\nu +
\beta A^\nu - \partial_\mu \partial^\nu A^\mu\label{15}
\end{equation}
Since the covariant Maxwell's equations follow Lorenz conditions
i.e. $\partial_\mu A^\mu = 0$ therefore the last term in the R.H.S
turns out to be zero [11]. So equation (\ref{15}) becomes
\begin{equation}
(p_\mu p^\mu + m^2 + \alpha l^2 p^4) A^\nu = \frac{4\pi}{c} j_\nu +
\beta A^\nu\label{16}
\end{equation}
Differentiating equation (\ref{16}) w.r.t $\partial_\nu$, we get
\begin{equation}
(p_\mu p^\mu + m^2 + \alpha l^2 p^4) \partial_\nu A^\nu =
\frac{4\pi}{c} \partial_\nu (j^\nu + \beta A^\nu)\label{17}
\end{equation}
Substituting equation (\ref{1}) in (\ref{17}) we get,
\begin{equation}
(E^2 - p^2 + m^2 + \alpha l^2 p^4) \partial_\nu A^\nu =
\frac{4\pi}{c}
\partial_\nu (j^\nu + \beta A^\nu )\label{18}
\end{equation}
We can see that equation (\ref{18}) is similar to conditions in
equation (\ref{15}). So under Lorenz conditions equation (\ref{18})
can be written as
\begin{equation}
(E^2 - p^2 + m^2 + \alpha l^2 p^4) \partial_\nu A^\nu = 0\label{19}
\end{equation}
which implies that
\begin{equation}
\frac{4\pi}{c} \partial_\nu (j^\nu + \beta A^\nu) = 0\label{20}
\end{equation}
We can see that equation (\ref{20}) gives some sort of an equation
of continuity and also this continuity is different from the usual
electromagnetic theory. [10]\\
Let us now try to find out the solutions in modified Maxwell's
equations under the condition that the LHS of equations (\ref{16})
LHS turns out to be zero.\\
Then we get,
\begin{equation}
(E^2 - p^2 + m^2 + \alpha l^2 p^4) A^\nu = 0\label{21}
\end{equation}
We consider plane wave solution for equation (\ref{21})
\begin{equation}
A_\mu = exp (\mp \imath k x) \epsilon_\mu (k)\label{22}
\end{equation}
Here $\epsilon_\mu (k)$  is the polarization vector and $k$ is the
wave vector.
\section{Results and Discussions}
The general solutions are given by    [9,11]
\begin{equation}
k^2 = m^2 + \alpha l^2 p^4, k^\nu \epsilon_\nu (k) = \frac{1}{2}
K^{\dot{A}B} \epsilon_{\dot{A}B} (K) = 0\label{23}
\end{equation}
where
$$K_{\dot{A}B} = k^\mu \sigma_{\mu ,\dot{A}B} = \left[\begin{array}{ll}
k^0 + k^3 \quad k^1 + \imath k^2\\
k^1 - \imath k^2 \quad k^0 - k^3\end{array}\right] \mbox{is \, a}\,
2 \times 2 \mbox{matrix}$$ For $m \ne 0$  there are three linearly
independent, space like polarization vectors $\epsilon^\mu_\imath
(k)$which are usually orthonormalized according to
$$\epsilon^\mu_\imath (k) \epsilon^*_{j,\mu} (k) = \frac{1}{2}
\epsilon_{\imath ,\dot{A}B} (k) \epsilon^{*\dot{A}B}_j (k) =
-\delta_{\imath j} , \imath , j = 0, \pm$$ where
$$\epsilon^{*\dot{A}B} (k) = \epsilon^*_\mu \sigma^{\mu ,\dot{A}B}$$
$$\epsilon_{+\dot{A}B} (K) = \sqrt{2}n_{2,\dot{A}} n_{1,B}, \epsilon_{-\dot{A}B} (K) =
\sqrt{2} n_{1,\dot{A}} n_{2,B},$$
\begin{equation}
\epsilon_{0,\dot{A}B} (K) = \frac{1}{m + \alpha lp^2}
(\kappa_{1,\dot{A}} \kappa_{1,B} -
\kappa_{2\dot{A}}\kappa_{2B})\label{24}
\end{equation}
where $n_{\imath \dot{A}}$ are the eigenvectors and $\kappa_{\imath
\dot{A}}$ are the wave vectors
$$n_{1,\dot{A}} = \left[\begin{array}{ll}
e^{-\imath \theta} cos \frac{\theta}{2}\\
sin \frac{\theta}{2}\end{array}\right], n_{1,\dot{A}} =
\left[\begin{array}{ll} sin \frac{\theta}{2}\\
-e^{-\imath \theta} cos \frac{\theta}{2}\end{array}\right] \quad
(See Appendix 1)$$ and $\kappa$ is defined as $\kappa_{\imath
,\dot{A}} = \sqrt{\lambda_\imath}
n_{\imath,\dot{A}}$  (See appendix 1) where $\lambda_1$  is the eigenvalue.\\
Now from the conjugate polarization vectors for equation (\ref{24})
are given by
$$\epsilon^*_{+\dot{A}B} (K) = \sqrt{2} n_{1,\dot{A}} n_{2,B}, \epsilon^*_{-\dot{A}B}
(K) = \sqrt{2} n_{2,\dot{A}} n_{1,B},$$
\begin{equation}
\epsilon^*_{0,\dot{A}B} (K) = \frac{1}{m + \alpha l p^2}
(\kappa_{1,\dot{A}} \kappa_{1,B} - \kappa_{2,\dot{A}}
\kappa_{2,B})\label{25}
\end{equation}
This $\epsilon^*_(\imath)$ is the conjugate polarization vector of
$\epsilon_(\imath)$ given in equation (\ref{24}) and (\ref{25}). Now
from the above equations we obtain the relations
\begin{equation}
\epsilon_{\imath ,\dot{A}B} (k) = \epsilon_{-\imath ,B\dot{A}} (k) =
\epsilon^*_{-1,\dot{A}B} (k) = \epsilon^*_{1,B\dot{A}} (k)\label{26}
\end{equation}
Finally we got the spinors for the modified Proca equation. The
three polarization vectors are converted to spinors. We neglect
$m^2$ in equation (\ref{9}), but not m in equation (\ref{24}) since
as we will see $m$ is small.\\
If we consider a case where the mass of the particle $m \to 0$ in
the equations (\ref{24}) and (\ref{25}). Then we can see that we
retain all the three equations which could not be possible from the
usual Proca equation.
$$\epsilon_{+\dot{A}B} (K) = \sqrt{2} n_{2,\dot{A}} n_{1B}, \epsilon_{-\dot{A}B} (K) =
\sqrt{2} n_{1,\dot{A}} n_{2,B},$$
\begin{equation}
\epsilon_{0,\dot{A}B} (K) = \frac{1}{\alpha l p^2}
(\kappa_{1,\dot{A}}\kappa_{1,B} - \kappa_{2,\dot{A}}
\kappa_{2,B})\label{27}
\end{equation}
And conjugate polarization vector are given by
$$\epsilon^*_{+\dot{A}B} (K) = \sqrt{2} n_{1,\dot{A}} n_{2,B}, \epsilon^*_{-\dot{A}B}
(K) = \sqrt{2} n_{w,\dot{A}} n_{1,B},$$
\begin{equation}
\epsilon^*_{0,\dot{A}B} (K) = \frac{1}{\alpha l p^2}
(\kappa_{1,\dot{A}} \kappa_{1,B} - \kappa_{2,\dot{A}}
\kappa_{2,B})\label{28}
\end{equation}
Now we can see that all three polarized states are retrieved with a
term $\alpha l p^2$  which equal to the mass of the photon, let's
see what will be the mass for a suitable wavelength.\\
Now consider the term $\alpha l p^2$ which is nothing but equivalent
to mass of photon $m_p$
\begin{equation}
m_P = \alpha l p^2\label{29}
\end{equation}
\begin{equation}
m_P = \alpha l \frac{(\hbar \omega)^2}{c^2}\label{30}
\end{equation}
Here we take the $\omega$ to be in the order of X-rays range
(i.e.$\omega = 10^{20} s^{-1}$ and $\hbar = 1.054 \times 10^{-27}
erg-s)$ and we can see below that
$$m_P = l_P \frac{(1.054 \times 10^{-27} \times 10^{20})^2}{(3
\times 10^{10})^2}$$
$$m_P \sim (10^{-65}) gms$$
taking
$$l_P \sim 10^{-33}cm$$
This is the same mass deduced by one of the authors (Sidharth) some
years ago [6] and recently conformed in the observation by the MAGIC
team. It is within the best experimental limit set for the photon
mass. [7, 8]
\section{Modified Weyl's Equation}
These Weyl equations were written for a neutrino initially which was
consider to be a mass less particle, but later experimental
observation showed that neutrino too have a mass. We will see now
how Weyls equation is modified with the inclusion of the extra
term.\\
Now lets us consider Weyl's equation with that extra term in the S-S
Hamiltonian
\begin{equation}
[p_0 - \vec{\bar{\sigma}} \cdot p + \alpha l \gamma^5 p^2]
\tilde{\phi}_R (p) = 0\label{31}
\end{equation}
and the other is
\begin{equation}
[p_0 + \vec{\bar{\sigma}} \cdot p + \alpha l \gamma^5 p^2]
\tilde{\phi}_L (p) = 0\label{32}
\end{equation}
We here we consider the $\vec{\bar{\sigma}}$  to be a $4 \times 4$
matrix and $\phi (p)$ are called the Weyl spinors which are also $4
\times 4$ , this is due to that extra term. Here after a bit of
algebra we will have two equations which represented as Modified
Weyl equation.\\
Now consider the equations (\ref{22}).Here we will take the complex
conjugate of (\ref{23}) then we get
\begin{equation}
\tilde{\phi}^*_L (p) [p^*_0 + \vec{\bar{\sigma}}^* \cdot p + \alpha
l (\gamma^5)^* p^2] = 0\label{33}
\end{equation}
Here we multiply (\ref{31}) and (\ref{32}). The term
$\tilde{\phi}^*_L (p)$ and $\tilde{\phi}_R (p)$ which represents a
spinors will have one row matrix and other column matrix multiplying
this two will give a scalar quantities which can be take out.
\begin{equation}
[p_0 - \vec{\bar{\sigma}} \cdot p + \alpha l \gamma^5 p^2]
\tilde{\phi}_R (p) \tilde{\phi}^*_L (p) [p^*_0 + \vec{\bar{\sigma}}
\cdot p + \alpha l (\gamma^5)^* p^2] = 0\label{34}
\end{equation}
\begin{equation}
[p_0 - \vec{\bar{\sigma}} \cdot p + \alpha l \gamma^5 p^2] [p^*_0 +
\vec{\bar{\sigma}} \cdot p + \alpha l (\gamma^5)^* p^2]
\tilde{\phi}_R (p) \tilde{\phi}^*_L (p) = 0\label{35}
\end{equation}
\begin{equation}
[p^2_0 - p^2 + \alpha l^2 p^4] \tilde{\phi}_R (p) \tilde{\phi}^*_L
(p) = 0\label{36}
\end{equation}
\begin{equation}
[E^2 - p^2 + \alpha l^2 p^4] \tilde{\phi}_R (p) \tilde{\phi}^*_L (p)
= 0\label{37}
\end{equation}
These represent a scalar bosons and a scalar wave function
$\tilde{\phi}_R (p) \tilde{\phi}^*_L (p)$. [11]
\section{Conclusion}
From modified Maxwell's equation we can see that we are getting a
new type of equation of continuity. And also we have got the value
of photon mass which turns out to be $m_P \sim (10^{-65}gms$ and
also a new scalar boson.
\newpage
\begin{center}
{\bf \large{APPENDIX-1}[9]} \end{center} \begin{center} {\bf
\large{CONVERSION OF FOUR VECTORS INTO SPINORS}} \end{center}
Minkowski four-vectors belong to the representation $D(\frac{1}{2},
\frac{1}{2} = D (\frac{1}{2} , 0) \otimes D (0, \frac{1}{2})$ of the
Lorentz group. The transition of the usual form of a four-vector
$k^\mu = (k^0 , \vec{k})$ to the spinor
representation is provided by the matrices\\
where $\sigma$ are the Pauli spin matrices.\\
$$\sigma^{\mu , \dot{A}B} = (\sigma^0 , \sigma), \quad \quad
\sigma^\mu_{\dot{A}B} = (\sigma^0 , - \sigma ) \quad \quad (b1)$$
consisting of the two-dimensional unit matrix $\sigma^0$  and the
Pauli matrices $\sigma^a$. Each four vector $k^\mu$ is related to a
$2 \times 2$ matrix
$$K_{\dot{A}B} = k^\mu \partial_{\mu ,\dot{A}B} = \left[\begin{array}{ll}
k^0 + k^3 \quad k^1 + \imath k^2\\
k^1 - \imath k^2 \quad k^0 - k^3\end{array}\right] \quad \quad
(b2)$$ this is Hermitian if the components of $k^\mu$ are real. The
rules for dotting, undotting, raising, and lowering spinor indices
also apply to the indices of the   matrices; in particular, we have
$$\sigma^\mu_{\dot{A}B} = \sigma^{\mu ,\dot{C}D} \epsilon_{\dot{C}\dot{A}} \epsilon_{DB},
\quad \sigma^\mu_{\dot{A}B} = (\sigma^\mu_{\dot{A}B})^* \quad \quad
(b3)$$ where $\epsilon = \imath \sigma^2$ and
$\epsilon_{\dot{C}\dot{A}} = \epsilon_{DB} =
\left[\begin{array}{ll} 0 \quad 1\\
-1 \quad 0\end{array}\right]$\\
We note that the coefficients of the transpose of a matrix $K^T$
read $K_{B\dot{A}}$ if the ones of $K$ are denoted by $K_{\dot{A}B}$
; i.e., transposing a matrix interchanges the spinor indices without
moving the position of the overdot. Thus the Hermiticity of the
matrices is expressed by
$$\sigma^{\mu ,\dot{A}B} = \sigma^{\mu ,B\dot{A}}, \quad \sigma^\mu_{\dot{A}B} =
\sigma^\mu_{B\dot{A}} \quad \quad (b4)$$ The $\sigma$  matrices
obeys the important relations
$$\sigma^\mu_{\dot{A}B} \sigma^{\nu ,\dot{A}B} = 2g^{\mu \nu}, \quad
\sigma^\mu_{\dot{A}B} \sigma^{\nu ,\dot{A}\dot{C}} +
\sigma^\nu_{\dot{A}B} \sigma^{\mu ,\dot{A}\dot{C}} = 2g^{\mu \nu}
\delta^{\dot{C}}B$$
$$\quad \quad \quad \sigma^\mu_{\dot{A}B} \sigma_{\mu ,\dot{C}D} =
2\epsilon_{\dot{A}\dot{C}} \epsilon_{BD} \quad \quad (b5)$$ The
first of these relations translates the Minkowski inner product of 2
four-vector $k^\mu$ and $p^\mu$ into
$$2k \cdot p = k_\mu 2g^{\mu \nu} p_\nu = k_\mu \sigma^\mu_{\dot{A}B}
\sigma^{\nu ,\dot{A}B} p_\nu = K_{\dot{A}B} p^{\dot{A}B}. \quad
\quad (b6)$$ where
$$p^{\dot{A}B} = \sigma^{\sigma \dot{A}B} p_\nu = \left[\begin{array}{ll}
p^0 + p^3 \quad p^1 + \imath p^2\\
p^1 - \imath p^2 \quad p^0 - p^3\end{array}\right]$$ and the second
term one implies
$$K_{\dot{A}B} K^{\dot{A}\dot{C}} = k^2 \delta^{\dot{C}}_B \quad \quad (b7)$$
In order to reduce terms involving a matrix $K_{\dot{A}B}$ to spinor
products, it is necessary to express $K_{\dot{A}B}$  in terms of
spinors. For a real four-vector, the matrix $K_{\dot{A}B}$  is
Hermitian and can be decomposed into its eigenvectors $n_{\imath
,\dot{A}} (\imath = 1,2)$ and eigenvalues $\lambda_\imath$
$$K_{\dot{A}B} = \sum_{\imath = 1,2} \lambda_\imath n_{\imath ,\dot{A}} n_{\imath ,B}
, \lambda_{\imath ,2} = k^0 \pm |k|,$$
$$n_{1,\dot{A}} = \left[\begin{array}{ll}
e^{-\imath \theta} cos \frac{\theta}{2}\\
sin \frac{\theta}{2}\end{array}\right], \quad \quad n_{2,\dot{A}} =
\left[\begin{array}{ll} sin \frac{\theta}{2}\\
-e^{-\imath \theta} cos \frac{\theta}{2}\end{array}\right] \quad
\quad (b8)$$ where $\theta$  and $\phi$ denoted the polar and
azimuthal angle of $k = |k|e$,respectively
$$\quad \quad \quad e = \left[\begin{array}{ll}
cos \phi sin \theta\\
sin \phi sin \theta\\
cos \theta\end{array}\right] \quad \quad (b9)$$ For time like
vectors $k^2 > 0$, it is often convenient to include the eigenvalues
of $\lambda_1$ in the normalization of the eigenvectors resulting in
$$\kappa_{\dot{A}B} = \sum_{\imath = 1,2} \kappa_{\imath ,\dot{A}} \kappa_{\imath ,B}, \quad \quad \kappa_{\imath ,\dot{A}}
= \sqrt{\lambda_\imath} n_{\imath ,\dot{A}} \quad \quad (b10)$$ The
phases $n_{1,\dot{A}}$ are chosen such that the orthonormality
relations read
$$<n_\imath n_\imath > = 0, \quad \quad <n_2n_1> = <n_1n_2> = +1
\quad \quad (b11)$$ The special case of a light like vector $(k^2 =
0)$ is of particular importance. In this case the eigenvalue
$\lambda_2$ of Equation $(b8)$ vanishes, and the matrix
$K_{\dot{A}B}$ factorizes into a single product of two spinors
$$K_{\dot{A}B} = k^\mu \sigma_{\mu ,{\dot{A}}B} = k_{\dot{A}} k_B, \quad \quad (b12)$$
$$k_{\dot{A}} = \sqrt{2k^0} n_{1,{\dot{A}}} = \sqrt{2k^0} \left[\begin{array}{ll}
e^{-\imath \phi} cos \frac{\theta}{2}\\
sin \frac{\theta}{2}\end{array}\right] \quad \quad (b12)$$ In this
context, $k_{\dot{A}}$  is called a momentum spinor.\\
Finally, we remark that the decomposition $(b8)$ is a very
convenient, but not unique, possibility to express a fourvector
$k^\mu$ with $k^2 \ne 0$ in terms of WvdW spinors. Any splitting of
$k^\mu$ into two lightlike four-vectors yields decomposition into
spinors, since light like vectors factorize, as seen above.\\
For instance, choosing an arbitrary light like fourvector $a^\mu
(a^2 \ne 0)$ with $a \cdot k > 0$ and defining
$$a = \frac{k^2}{2a \cdot k}, \quad \quad b^\mu = k^\mu - a a^\mu
\quad \quad (b13)$$ yields a possible decomposition $k^\mu = a a^\mu
+ b^\mu$. In terms of WvdW spinor, this correspond to an arbitrarily
chosen spinor $a_{\dot{A}}$ with $K_{{\dot{C}}D} a^{\dot{C}} a^D >
0$, leading to the decomposition
$$K_{{\dot{A}}B} = a a_{\dot{A}} a_B + b_{\dot{A}}b_B$$
with
$$b_{\dot{A}} = - \frac{K_{B{\dot{A}}} a^B}{\sqrt{K_{{\dot{C}}D}a^{\dot{C}} a^D}}, \quad a =
\frac{k^2}{k_{{\dot{C}}D} a^{\dot{C}} a^D} \quad \quad (b14)$$

\end{document}